\documentclass[nojss]{jss}

\usepackage{thumbpdf,lmodern}

\newcommand{\fct}[2][]{\code{#2(#1)}}

\usepackage{amsmath}
\usepackage{amssymb}
\usepackage{algpseudocode,algorithm,algorithmicx} 
\usepackage{natbib}

\DeclareMathOperator*{\argmin}{arg\,min}

\author{Anton Alyakin\\Johns Hopkins University
  \And Yichen Qin\\University of Cincinnati
  \And Carey E. Priebe\\Johns Hopkins University}
\Plainauthor{Anton Alyakin, Yichen Qin, Carey E. Priebe}

\title{\pkg{LqRT}: Robust Hypothesis Testing of Location Parameters using 
Lq-Likelihood-Ratio-Type Test in \proglang{Python}}
\Plaintitle{LqRT: Robust Hypothesis Testing of Location Parameters using 
	Lq-Likelihood-Ratio-Type Test in Python}
\Shorttitle{LqRT: Robust Lq-Likelihood-Ratio-Type Test}

\Abstract{
  A $t$-test is considered a standard procedure for inference on population
  means and is widely used in scientific discovery.
  However, as a special case of a likelihood-ratio test, $t$-test often shows
  drastic performance degradation due to the deviations from its hard-to-verify
  distributional assumptions.
  Alternatively, in this article, we propose a new two-sample
  L$q$-likelihood-ratio-type test (L$q$RT) along with an easy-to-use
  \proglang{Python} package for implementation.
  L$q$RT preserves high power when the distributional assumption is violated, 
  and maintains the satisfactory performance when the assumption is valid.
  As numerical studies suggest, L$q$RT dominates many other robust tests in
  power, such as Wilcoxon test and sign test, while maintaining a valid size.
  To the extent that the robustness of the Wilcoxon test (minimum asymptotic 
  relative efficiency (ARE) of the Wilcoxon test vs the $t$-test is 0.864) 
  suggests that the Wilcoxon test should be the default test of choice (rather 
  than ``use Wilcoxon if there is evidence of non-normality,'' the default 
  position should be ``use Wilcoxon unless there is good reason to believe the 
  normality assumption''), the results in this article suggest that the 
  L$q$RT is potentially the new default go-to test for practitioners.
}

\Plainkeywords{Lq-likelihood, Testing, Robustness, Python}

\Address{
  Anton Alyakin\\
  Department of Applied Mathematics and Statistics\\
  Johns Hopkins University\\
  3400 Charles St.\\
  Clark Hall, 319A\\
  21218 Baltimore, MD\\
  USA\\
  E-mail: \email{aalyaki1@jhu.edu}\\
  URL: \url{https://alyakin314.github.io} \\
  \\
  Yichen Qin\\
  Department of Operations, Business Analytics, and Information Systems\\
  University of Cincinnati\\
  2906 Woodside Dr.\\
  45208 Cincinnati, OH\\
  USA\\
  E-mail: \email{qinyn@ucmail.uc.edu}\\
  URL: \url{https://sites.google.com/view/yichenqin/home} \\
  \\
  Carey Priebe\\
  Department of Applied Mathematics and Statistics\\
  Johns Hopkins University\\
  3400 Charles St.\\
  Clark Hall, 301D\\
  21218 Baltimore, MD\\
  USA\\
  E-mail: \email{cep@jhu.edu}\\
  URL: \url{https://www.ams.jhu.edu/~priebe/}
}
\begin{document}

\section[Introduction]{Introduction} \label{sec:intro}
Classical testing procedures, such as $t$-test, claim its optimal performance
and its highest power while relying heavily on the assumptions that are neither
verifiable nor actually hold in many real data settings.
For example, datasets often include outliers, observations that deviate
significantly from other elements of the sample, and are unlikely under the
assumed underlying distribution \citep{Grubbs:1969}.
Outliers' presence may be due to human error, instrument malfunction or the
complexity of the true data-generating process compared to the model
\citep{Qin+Priebe:2017}.
In the presense of outliers, the validity of the classical test is undermined 
and its power is severely degraded.
Therefore, it is important to develop statistical hypothesis testing procedures 
that are not significantly affected by their presence.

An ideal robust method should show nearly optimal performance 
when outliers are present, or more generally, under the deviations from 
distributional assumptions \citep{Hampel:2011}, and maintains satisfactory 
performance when assumptions are valid.
For estimation, such examples include median or trimmed mean as measures of 
central 
tendency.
For hypothesis testing, there has been relatively less investigation.

Historically, robust testing procedures are predominantly non-parametric.
The first significance test used was a robust test, when in 1710 John Arbuthnot
used a sign test to conclude that the probability of male and female births are
not exactly equal \citep[][pp. 157-176]{Conover:1999}
Other robust non-parametric testing procedures include Wilcoxon signed-rank test
\citep{Wilcoxon:1945} and Wilcoxon rank-sum test, also known as Mann-Whitney $U$
test \citep{Mann+Whitney:1947}.
Parametric robust tests are less common because they naturally depend on the 
parametric distribution assumptions.
The most notable example of such is the Huber Ratio Test \citep{Huber:1965}.

Recently, \cite{Qin+Priebe:2017} have proposed a robustified likelihood ratio 
test by leveraging the idea of L$q$-likelihood \citep{Ferrari+Yang:2010}, which 
they call L$q$-likelihood-ratio-type test (L$q$RT).
By assigning observations different weights according to their likelihoods 
under the assumed model, 
L$q$RT is able to attain powers similar to parametric tests when model 
assumptions are valid, 
and maintain satisfactory powers similar to or higher than nonparametric 
tests when assumptions are violated, leading to a uniform domination over 
Wilcoxon test and sign test. 

In the literature, it has been shown that the minimum asymptotic relative 
efficiency (ARE) of 
the Wilcoxon test vs the $t$-test is 0.864 \citep{Hodges+Lehmann:1956}, 
which advocates that the Wilcoxon test should be the default 
test of choice as opposed to $t$-test, 
and that the practitioners should ``use Wilcoxon unless there 
is good reason to believe the normality assumption,'' 
as opposed to ``use Wilcoxon if there is evidence of non-normality.''
The desirable results of L$q$RT suggest that L$q$RT is potentially the new 
default go-to test for practitioners.

However, the original L$q$RT paper focuses mostly on the theoretical properties 
and is lack of easy implementation of the test for practitioners.
In addition, it also misses many important cases such as two-sample 
paired/unpaired tests for the equivalence of locations, which are frequently 
encountered in scientific discovery.
To unleash the power of L$q$RT, in this work, we present the extension of 
L$q$RT to the two-sample case, propose a new approach to select the tuning 
parameter $q$, more importantly, provide \pkg{LqRT}, a \proglang{Python} 
package which implements the one-sample, the two-sample paired and the 
two-sample unpaired L$q$RT, so that practitioners can easily adopt the proposed 
test in their analysis.

\proglang{Python} \citep{Python} is an interpreted, high-level, general-purpose
programming language.
There are a variety of statistical significance tests implemented in
\proglang{Python}.
The scientific computing package \pkg{SciPy} \citep{SciPy} implements one-sample
$t$-test as \fct{scipy.stas.ttest\_1samp}, two-sample paired $t$-test as
\fct{scipy.stas.ttest\_rel}, and two-sample unpaired $t$-test as
\fct{scipy.stas.ttest\_ind}.
It also contains non-parametric Wilcoxon Rank-sum and Wilcoxon Signed-Rank
tests, implemented as \fct{scipy.stats.ranksums} and
\fct{scipy.stats.wilcoxon}, respectively.
\pkg{LqRT} uses a syntax that is similar to that of \pkg{SciPy}.
There is no sign test in \pkg{SciPy}, but it is implemented as
\fct{statsmodels.stats.descriptivestats.sign\_test} in \pkg{statsmodels}
\citep{Statsmodels}.  The proposed \pkg{LqRT} is a natural complement to the 
existing toolbox.  As shown in our simulation studies, \pkg{LqRT} offers much 
improved numerical performance while requires little modification of analysts' 
existing code.

The rest of the paper is organized as follows.
In Section \ref{sec:methods} we present the L$q$RT procedure.
Specifically, we first review the foundations of the 
L$q$-likelihood, and then present the general form of L$q$RT. 
We further extend it to one-sample, paired two-sample and unpaired two-sample 
tests. 
Finally, we discuss the method to estimate the $p$-value and the tuning 
parameter $q$.
We present the \proglang{Python} package, \pkg{LqRT}, and its core functions 
and their instructions in Section \ref{sec:software}.
In Section \ref{sec:experiments} we demonstrate the performance of
\pkg{LqRT} compared to other tests implemented in \proglang{Python}.
We conclude and discuss our results in Section \ref{sec:conclusion}.
Additional information such as the pseudo-codes are included in the Appendix 
\ref{app:algorithms}.

\section[Methodology]{Lq-likelihood-ratio-type test} \label{sec:methods}

\subsection{Preliminaries} \label{sec:mlqe}

Let $\boldsymbol{x} = \left[x_1, \hdots, x_n \right]^T$ denote a sample of
independent observations from a hypothesized distribution 
$f(\cdot|\boldsymbol{\theta})$ parameterized by $\boldsymbol{\theta}$.
The log-likelihood of this sample is $\sum_{i=1}^n \log ( f\left(x_i 
\vert \boldsymbol{\theta} \right))$, while the L$q$-likelihood 
\citep{Ferrari+Yang:2010} of the sample is 
defined as 
\begin{align*}
  \sum_{i=1}^n L_q \left(f \left(x_i \vert \boldsymbol{\theta} \right)\right),
\end{align*}
where $L_q(\cdot)$ is a q-deformed logarithm first introduced in
\cite{Tsallis:1994}.
Specifically, for any positive $u$,
\begin{align*}
  L_q(u)
  &=
    \begin{cases}
      \log (u) & \mathrm{if} \quad q=1 \\
      \frac{u^{1-q} - 1}{1 - q} & \mathrm{otherwise}
    \end{cases}.
\end{align*}
This function is equivalent to the Box-Cox transformation under a $\lambda = 1 -
q$ reparameterization \citep{Box+Cox:1964}.
Note that $L_q (u) \to \log (u)$ as $q \to 1$.  
Therefore L$q$-likelihood includes log-likelihood as speical case $q=1$.  
The maximum L$q$-likelihood estimation (ML$q$E) of $\boldsymbol{\theta}$, 
$\widehat{\boldsymbol{\theta}}$, can 
be naturally defined as the maximizer of L$q$-likelihood 
\citep{Ferrari+Yang:2010},
\begin{align*}
\widehat{\boldsymbol{\theta}}=\arg\max_{\boldsymbol{\theta}}\sum_{i=1}^n L_q 
\left(f 
\left(x_i 
\vert 
\boldsymbol{\theta} \right)\right)
\end{align*}

By solving for the root of the gradient of the L$q$-likelihood, 
ML$q$E satifies
\begin{align*}
  \boldsymbol{0}=\nabla_{\boldsymbol{\theta}}
  \sum_{i=1}^n L_q \left(x_i \vert \boldsymbol{\theta} \right) 
  \Bigg|_{\boldsymbol{\theta}=\hat{\boldsymbol{\theta}}}
  &=\sum_{i=1}^n 
    \frac{\nabla_{\boldsymbol{\theta}} 
  	f\left(x_i \vert \boldsymbol{\theta}\right)}
    {f\left(x_i \vert \boldsymbol{\theta}\right)}
    f\left(x_i \vert \boldsymbol{\theta}\right)^{1-q} 
    \Bigg|_{\boldsymbol{\theta}=\hat{\boldsymbol{\theta}}}
	= \sum_{i=1}^n
    \nabla_{\boldsymbol{\theta}} \log f\left(x_i \vert 
    \boldsymbol{\theta}\right) w_i 
    \Bigg|_{\boldsymbol{\theta}=\hat{\boldsymbol{\theta}}}
\end{align*}
where $w_i := f\left(x_i \vert \boldsymbol{\theta}\right)^{1-q}$.
Thus, it is a weighted version of the gradient of the log-likelihood where the
weights are likelihoods taken to the power of $1-q$.
Clearly, for $q = 1$, the weights are all $1$, so the ML$q$E coincides with the
MLE.
For $q < 1$, this reweighting allows to reduce the effect of potential outliers 
whose likelihood tend to be small.
From hereafter, we formalize the notion of an outlier to such points with small
likelihood.  
With the preliminaries introduced above, we are ready to introduce our proposed 
hypothesis testing procedure.

\subsection{Lq-likelihood-ratio-type test} \label{sec:lqrt}
Suppose we have a sample $\boldsymbol{x}$ with a hypothesized density $f(x_i |
\theta)$ and we are interested in testing the parameter $\theta$ of this 
density for $H_0: \theta \in \Theta_0$ against $H_1: \theta \in \Theta_1$.
Assume that this parameter $\theta$ is a location parameter of a symmetric
density $f$.
Then we define the L$q$-likelihood-ratio-type (L$q$RT) as follows:  The test 
statistic is
\begin{align*}
  D_q(\boldsymbol{x})
  &=  2 \sup_{\theta \in \Theta_0 \cup \Theta_1}
    \left\{ \sum_{i=1}^n L_q \left(f(x_i | \theta) \right) \right\}
    - 2 \sup_{\theta \in \Theta_0}
    \left\{ \sum_{i=1}^n L_q \left(f(x_i | \theta) \right) \right\},
\end{align*}
and we reject the null hypothesis for the large values of $D_q$.  
This test falls in the category of the likelihood-ratio-type test, defined in
\cite{Heritier+Ronchetti:1994}.
Under some regularity conditions, the asymptotic distribution of the test 
statistic is a $\chi^2$ for $q \in [0, 1]$.
For $q = 1$, L$q$RT coincides with likelihood ratio 
test (LRT).
For $q < 1$, L$q$RT has been shown to be more robust to the 
contamination than LRT, both theoretically and experimentally.

The robustness of L$q$RT comes from the fact the L$q$-likelihood function is 
less sensitive to outliers than the log-likelihood function, 
but remains approximately as sensitive to the rest of the data points as the 
log-likelihood function.
When the model assumptions is valid and data is clean, the L$q$RT behaves 
similarly to LRT.  
The test statistic follows a $\chi^2$ distribution under the null and a 
non-central $\chi^2$ distribution under the alternative.
When the data is contaminated with the gross error model, the test statistic of 
LRT becomes an inflated $\chi^2$ distribution (or inflated non-central 
$\chi^2$) where the inflation magnitude depends the level of contamination, 
leading to a much large overlap between the null and alternative distributions 
and power degradation.
On the other hand, the test statistic $D_q$ of L$q$RT is much less affected by 
the contamination with very little inflation, maintaining a small overlap 
between the null and alternative distributions and a relatively high power (see 
Figure 2 in \citet{Qin+Priebe:2017}).

The inflation of LRT in this case is mainly due to the extreme values of the 
log-likelihood of the outliers.
For example, an outlier $x_{outlier}$ under the assumed model $f(\cdot|\theta)$ 
usually takes a small likelihood, i.e., $f(x_{outlier}|\theta) \approx 0$. 
Taking the logarithm of a small likelihood, $\log f(x_{outlier}|\theta)$, leads 
to a large negative value, which inflates the test statistic.
Alternatively, for L$q$RT, L$q$ function is bounded from below (i.e., 
$L_q(u)>-1/(1-q)$) which offers the robustness and maintains a similar shape to 
$\log$ which offers the high power.
Because of this property, as shown later in the simulation, the L$q$RT is able 
to main high power similar to the parametric tests such as $t$-test when the 
model assumption is valid and maintain high power similar to the nonparametric 
tests such as Wilcoxon test and sign test.
Please refer to \cite{Qin+Priebe:2017} for a more detailed discussion of 
asymptotic properties of L$q$RT.

Note that the test statistic $D_q$ is the difference of the maximums of the 
L$q$-likelihoods under the null hypothesis and under the union of 
null and alternative hypothesis.  Therefore, to compute $D_q$, it is 
equivalently to obtain the ML$q$E under the null and alternative hypothesis
However, there is no closed form solution to the maximization of for the 
L$q$-likelihood function.
\cite{Ferrari:2008} suggests using an iterative mixture model EM-like 
re-weighting algorithm.
Similarly to a regular EM, this algorithm arises naturally from the fact that
the estimator would be easy to compute if the weights were known, whilst the
weights themselves are easy to compute as a function of the data and parameters
\citep{Murphy:2013}.  The general iterative reweighting algorithm for obtaining 
ML$q$E and the maximum of L$q$-likelihood is summarized 
in Algorithm \ref{alg:reweighting_general}.

\begin{algorithm}[t!]
	\caption{Iterative re-weighting algorithm for compute maximum of 
	L$q$-likelihood}
	\label{alg:reweighting_general}
	\begin{algorithmic}[1]
		\Function{MLqE}{$\boldsymbol{x}$, $q$}
		\State $n \gets $ \Call{Length}{$\boldsymbol{x}$}
		
		\Statex
		\Statex \quad \, \# Initialize Parameters with MLE
		\State $\hat{\theta}^{(0)} \gets MLE(\boldsymbol{x})$
		\Statex
		\Statex \quad \, \# Iterative Re-weighting
		\For{$s=1, 2, \hdots \ until\ convergence$}
		\Statex
		\Statex \qquad \quad \# Update Weights
		\For{$i \gets 1, 2, \hdots, n$}
		\State $w_i^{(s)} \gets  f ( x_i \vert \hat{\theta}^{(s-1)}) ^ {1 - q}$
		\EndFor
		\Statex
		\Statex \qquad \quad \# Estimate the parameters using updated weights
		\State $\hat{\theta}^{(s)} \gets \text{ the root of } 0 = \sum_{i=1}^n
		w_i^{(s)} \frac{\partial}{\partial \theta} \log f(x_i | \theta) $
		\Statex
		\EndFor
		\State \Return $\hat{\theta}^{(s)}$ as ML$q$E and $\sum_{i=1}^n
		L_q(f(x_i | \hat{\theta}^{(s)}))$ as the maximum of L$q$-likelihood.
		\EndFunction
	\end{algorithmic}
\end{algorithm}

Note that ML$q$E is generally neither consistent nor asymptotically unbiased.
There have been various methods proposed to modify the procedure in the way to
make it consistent for all problems.
This includes using sequences $q_n$ with a property $q_n \to 1$
\citep{Ferrari+Yang:2010} and correcting the inherent bias the estimator with an
additive term \citep{Qin+Priebe:2017}.
However, the estimation of the location parameter of a symmetric distribution, 
such as normal distribution,
is one of the few special cases for which the ML$q$E is consistent by itself.
We proceed with using not-bias corrected ML$q$E, since there is no bias to
correct for our parameter of interest, but we draw the readers' attention to the
fact that an asymptotically unbiased estimator for $\sigma^2$ can be obtained
by modifying the ML$q$E by a factor of $q$, $\hat{\sigma}_{corrected}^2 = q
\hat{\sigma}^2$

In our work we are mostly concerned with the normal distribution, most 
frequently used case in practice.
Below we discuss in more details the three cases of L$q$RT with the normal 
distribution assumption.

\subsection{One-sample test}

Suppose we are interested in testing for the mean of a normal distribution, 
i.e., $H_0: \mu = \mu_0$ versus $H_1: \mu \neq \mu_0$.  
Let $f(\cdot \vert \mu, \sigma^2)$ represent the normal density. 
The test statistic of L$q$RT is
\begin{align}
  D_q(\boldsymbol{x})
  &= 2 \sup_{\substack{ \mu \in \mathbb{R} \\ \sigma^2 \in \mathbb{R}^+ } }
  \left\{ \sum_{i=1}^n L_q \left(f(x_i | \mu, \sigma^2) \right) \right\}
   - 2 \sup_{\sigma^2 \in \mathbb{R}^+}
    \left\{ \sum_{i=1}^n L_q \left(f(x_i | \mu_0, \sigma^2) \right) \right\}.
\end{align}
Note that L$q$RT becomes one-sample $t$-test when $q=1$,
so L$q$RT can be considered as a robustified version of $t$-test.
To obtain the ML$q$E of the mean of a normal distribution and compute $D_q$, 
we use a special case of iterative reweighting algorithm outlined in Algorithm 
\ref{alg:reweighting_regular} to optimize the parameters for 
the first term and a known-mean version (Algorithm 
\ref{alg:reweighting_known_mean}) for the second term.
More specifically, Algorithm \ref{alg:reweighting_regular} determines an MLqE 
for both the mean and the variance of a sample that comes from a normal 
distribution.  
Such a algorithm is only applicable if there are no restrictions on the
parameters.

\begin{algorithm}[t!]
	\caption{Iterative re-weighting algorithm for an MLqE of a one sample from a
		one-dimensional normal with no additional constraints}
	\label{alg:reweighting_regular}
	\begin{algorithmic}[1]
		\Function{MLqE-Normal}{$\boldsymbol{x}$, $q$}
		\State $n \gets $ \Call{Length}{$\boldsymbol{x}$}
		
		\Statex
		\Statex \quad \, \# Initialize Parameters with MLE
		\State $\hat{\mu}^{(0)} \gets n^{-1} \sum_{i=1}^n x_i$
		\State $\hat{\sigma}^{2^{(0)}} \gets
		n^{-1} \sum_{i=1}^n (x_i - \hat{\mu}^{(0)})^2$
		\Statex
		\Statex \quad \, \# Iterative Re-weighting
		\For{$k=1, 2, \hdots \ until\ convergence$}
		\Statex
		\Statex \qquad \quad \# Update Weights
		\For{$i \gets 1, 2, \hdots, n$}
		\State $w_i^{(s)} \gets \left( f \left( x_i \vert \hat{\mu}^{(s-1)},
		\hat{\sigma}^{2^{(s-1)}}\right) \right) ^ {1 - q}$
		\EndFor
		\Statex
		\Statex \qquad \quad \# Update Parameters
		\State $\hat{\mu}^{(s)} \gets  \frac{\sum_{i=1}^n x_i}{\sum_{i=1}^n 
			w_i}$
		\State $\hat{\sigma}^{2^{(s)}} \gets \frac{\sum_{i=1}^n w_i (x_i -
			\hat{\mu}^{(s)})^2}{\sum_{i=1}^n w_i}$
		\Statex
		\Statex \qquad \quad \# Clip Variances
		\If{$\hat{\sigma}^{2^{(i)}} < \epsilon$} \label{clipping}
		\State $\hat{\sigma}^{2^{(i)}} \gets \epsilon$
		\EndIf
		\EndFor
		\State \Return $\hat{\mu}^{(s)}$, $\hat{\sigma}^{2^{(s)}}$
		\EndFunction
	\end{algorithmic}
\end{algorithm}

The clipping of the variance in step \ref{clipping} of Algorithm
\ref{alg:reweighting_regular} is required in the implementation to avoid the
division by $0$ in the cases when the variance shrinks down unceasingly.
The $\epsilon$ is usually chosen to be some very small number, for example, a 
numerical precision of the 64-bit floating point numbers in \proglang{Python}.

The shrinkage of the variance to $0$ in the ML$q$E is a similar issue to the
non-existence of the global MLE in the mixtures of Gaussians.
In both of these situations, the estimating distribution becomes centered
exactly at one of the data points, and the variance is allowed to decrease,
which explodes the objective function, the likelihood in the mixtures of
guassians or the L$q$-likelihood in the Gaussian ML$q$E.
The difference comes from the fact that in the mixtures case, this effect does
not simultaneously affect the likelihood of all other points because the other
components become responsible for them, whereas in ML$q$E, the weights of all
points that are not centered exactly at the mean shrink as variance shrinks.

When performing the numerical experiments, we have observed this phenomenon
predominantly in the bootstrapped samples used to estimated the p-value.
Specifically, this effect tends to occur in samples where one observation is
repeated a significant number of times.
To give a sense of a scale, only about $0.2\%$ of the bootstrapped samples
converged to a degenerate distribution when the dataset size was 100.
We have observed the shrinkage of the variance to $0$ on the actual samples,
as opposed to the ones bootstrapped by the testing procedure, but the
occurrences of this were extremely sparse.

\subsection{Two-sample paired test}
Suppose that we have two samples
$\boldsymbol{x} = \left[x_1, \hdots, x_n \right]^T$ and $\boldsymbol{y} = 
\left[y_1, \hdots, y_n \right]^T$, 
and the samples $(x_i, y_i)$ are paired.
For example, they correspond to the observations of the same patient in the
beginning and in the end of some longitudinal study.
The hypothesis to be tested is $H_0: \mu_x = \mu_y$ against $H_1: \mu_x \neq
\mu_y$.

Similarly to the one sample test above, we can define a set of new variables
$\boldsymbol{z} = \left[z_1, \hdots, z_n \right]$, such that $z_i = y_i - x_i$,
and then perform a one-sampled test for $H_0: \mu_z = 0$ against
$H_1: \mu_z \neq 0$, using the single sample procedure described above.

\subsection{Two-sample unpaired test}
Lastly, suppose we have two samples $\boldsymbol{x} = \left[x_1, \hdots, x_n
\right]^T$ and $\boldsymbol{y} = \left[y_1, \hdots, y_m \right]^T$, not
necessarily with the same size.
The first sample comes from a normal distribution with density
$f(x_i | \mu_x, \sigma_x^2)$
and the second from a normal distribution with a density
$f(y_i | \mu_y, \sigma_y^2)$.
We, once again, want to test for $H_0: \mu_x = \mu_y$ against $H_1: \mu_x \neq
\mu_y$.

It is possible to do so with and without the shared variance assumption.
In the former case we assume that $\sigma_x^2 = \sigma_y^2 = \sigma^2$, and
can use the test statistic given by
\begin{align}
  D_q(\boldsymbol{x}, \boldsymbol{y})
  = & 2 \sup_{\substack{ \mu_x , \mu_y \in \mathbb{R} \\ \sigma^2 \in 
  \mathbb{R}^+ }}
  \left\{ \sum_{i=1}^n L_q \left(f(x_i | \mu_x, \sigma^2) \right)
  + \sum_{j=1}^m L_q \left(f(y_i | \mu_y, \sigma^2) \right) \right\}\nonumber\\
    &- 2 \sup_{\substack{ \mu \in \mathbb{R} \\ \sigma^2 \in 
    		\mathbb{R}^+ }}
      \left\{ \sum_{i=1}^n L_q \left(f(x_i | \mu, \sigma^2) \right)
      + \sum_{i=1}^m L_q \left(f(y_j | \mu, \sigma^2) \right) \right\}
\end{align}
For the first term, we can combine the two samples together and use the
regular reweighting algorithm (Algorithm \ref{alg:reweighting_regular})
to determine the optimal parameters.
For the second term, we have to use the re-weighting algorithm that estimates
one shared variance, but two different means for two samples (Algorithm
\ref{alg:reweighting_shared_variance}).
This test procedure corresponds to a more robust version of the classical
Student's $t$-test.

We can also make no assumption of the shared variances and obtain the test that
is instead more similar in spirit to the Welch's $t$-test, but is more robust.
Specifically, we use the test statistic given by
\begin{align}
  D_q(\boldsymbol{x}, \boldsymbol{y})
  = &2 \sup_{\substack{ \mu_x , \mu_y \in \mathbb{R} \\ \sigma_x^2, \sigma_y^2 
  \in \mathbb{R}^+ }}
  \left\{ \sum_{i=1}^n L_q \left(f(x_i | \mu_x, \sigma^2) \right)
  + \sum_{j=1}^m L_q \left(f(y_i | \mu_y, \sigma^2) \right) \right\}\nonumber\\
  	&- 2 \sup_{\substack{ \mu \in \mathbb{R} \\ \sigma_x^2, \sigma_y^2 \in 
  	\mathbb{R}^+ }}
      \left\{ \sum_{i=1}^n L_q \left(f(x_i | \mu, \sigma^2) \right)
      + \sum_{i=1}^m L_q \left(f(y_j | \mu, \sigma^2) \right) \right\}
\end{align}
In order to optimize the first term, we use a constrained version of the
re-weighting algorithm that estimates one shared mean for two samples, but two
differnt variances (Algorithm \ref{alg:reweighting_shared_mean}).
For the second term, there are no shared parameters, so it is sufficient to
simply use the regular one-sampled re-weighting algorithm (Algorithm
\ref{alg:reweighting_regular}) on each of the two samples individually.

Modified versions of this algorithm that handle the case of two samples with a
shared mean or variance, or the case when the mean is known a priori, are
presented in the Appendix \ref{app:algorithms}.

\subsection{Estimating $p$-values} \label{sec:critical_value}
We have discussed how to compute the test statistic $D_q$, but so far made no
statement on how to determine whether it falls inside the rejection region.
In fact, the test statistic follows a scaled chi-square distribution.
In practice, instead of obtaining the null distribution, we estimate the 
p-values.

In our implementation of the one-sample test, we estimate the $p$-value
using the bootstrap procedure for the location parameter test provided in
\cite{Qin+Priebe:2017}.
This procedure relies on transforming the original sample to be centered around
the null hypothesis mean and then bootstrapping the distribution of $D_q$
under the null.
The exact algorithm for doing so is outlined in Algorithm \ref{alg:bootstrap_p}.

\begin{algorithm}[t!]
  \caption{Bootsrap procedure to estimate the $p$-value for the one-sample
    L$q$RT.}
  \label{alg:bootstrap_p}
  \begin{algorithmic}[1]
    \Function{p-Value-1Sample}{$\boldsymbol{x}$, $u$, $q$, $B$}
    \State $\hat{u}, \hat{\sigma}^2 \gets $\Call{MLqE-Normal}{$\boldsymbol{x}$, $q$}
    \State $\boldsymbol{x}' \gets \left[x_1 - \hat{u} + u, \hdots, x_1 -
      \hat{u} + u \right]^T$
    \For{$b = 1, \hdots, B$}
    \State $\boldsymbol{x}_b' \gets $ \Call{Resample}{$\boldsymbol{x}'$} 
    \State $D_q^b \gets D_q(\boldsymbol{x}_b')$
    \EndFor
    \State $D_q \gets D_q(\boldsymbol{x})$
    \State $\widehat{p} \gets$ quantile of $D_q^b$s that are greater
    than $D_q$
    \State \Return $\widehat{p}$
    \EndFunction
  \end{algorithmic}
\end{algorithm}

\begin{algorithm}[t!]
  \caption{Bootsrap procedure to estimate the $p$-value for the two-sample
    unpaired L$q$RT.}
  \label{alg:bootstrap_p_two_sample}
  \begin{algorithmic}[1]
    \Function{p-Value-2Sample-unpaired}{$\boldsymbol{x}$, $\boldsymbol{y}$, $q$,
      $B$}
    \State $\hat{u}_x, \hat{\sigma}_x^2 \gets $
    \Call{MLqE-Normal}{$\boldsymbol{x}$, $q$}
    \State $\hat{u}_y, \hat{\sigma}_y^2 \gets $
    \Call{MLqE-Normal}{$\boldsymbol{y}$, $q$}
    \State $\boldsymbol{x}' \gets \left[x_1 - \hat{u}_x, \hdots, x_1 -
      \hat{u}_x \right]^T$
    \State $\boldsymbol{y}' \gets \left[y_1 - \hat{u}_y, \hdots, x_1 -
      \hat{u}_y \right]^T$
    \For{$b = 1, \hdots, B$}
    \State $\boldsymbol{x}_b' \gets $ \Call{Resample}{$\boldsymbol{x}'$} 
    \State $\boldsymbol{y}_b' \gets $ \Call{Resample}{$\boldsymbol{y}'$} 
    \State $D_q^b \gets D_q(\boldsymbol{x}_b', \boldsymbol{y}_b')$
    \EndFor
    \State $D_q \gets D_q(\boldsymbol{x}, \boldsymbol{y})$
    \State $\widehat{p} \gets$ quantile of $D_q^b$s that are greater
    than $D_q$
    \State \Return $\widehat{p}$
    \EndFunction
  \end{algorithmic}
\end{algorithm}

For the two-sample unpaired test we propose a procedure that is very similar to
the Algorithm \ref{alg:bootstrap_p}, except both samples are centered around
$0$.
This is accomplished by subtracting a robustly estimated means from the
respective samples.
The procedure is outlined in the Algorithm \ref{alg:bootstrap_p_two_sample}.
This allows to bootstrap data from the null hypothesis under which both samples
have the same means.

\subsection{Selecting $q$} \label{sec:select_q}
The tuning parameter $q$ is important in our testing procedure because it 
controls the trade of between power and robustness, 
and it is generally not known a priori. 
In our approach, we select $q$ by minimizing the trace of the
asymptotic covariance matrix of $\widehat{\boldsymbol{\theta}}_q$.
In a one-dimensional Gaussian case for a single sample this implies selecting
\begin{align*}
  \hat{q}
  &= \argmin_q \left\{ \hat{V}_q(\hat{\mu}_{q}) \right\} \\
  &= \argmin_q \left\{ a_q b_q a_q \right\}
\end{align*}
where
\begin{align*}
  a_q &= \left( \frac{1}{n} \sum_{i=1}^n
        \frac{\partial^2} {\partial \mu^2 } L_q \left(f (x_i \vert \mu, 
        \sigma^2) \right)\middle|_{\mu = \hat{\mu}_q, \sigma^2 = 
        \hat{\sigma}_q^2}
        \right)^{-1}
\end{align*}
and
\begin{align*}
  b_q &= \frac{1}{n} \sum_{i=1}^n \left( 
        \frac{\partial} {\partial \mu } L_q \left(f (x_i \vert \mu, \sigma^2) 
        \right) \middle|_{\mu = \hat{\mu}_q, \sigma^2 = \hat{\sigma}_q^2}
        \right)^{2}
\end{align*}
The $\hat{\mu}_q$ and $\hat{\sigma}_q^2$ are ML$q$E estimates over the whole
hypothesis space.
For a two-sample unpaired case, we select $q$ in a similar fashion
\begin{align*}
  \hat{q}
  &= \argmin_q \left\{ \hat{V}_q(\hat{\mu}_{y,q})
    + \hat{V}_q(\hat{\mu}_{x,q}) \right\}
\end{align*}
and the respective empirical variances are computed in a similar way to the
one-dimensional case.

In our implementation, we use a grid search over the values of $q \in [0.5,
1.0]$ with an interval of $0.01$ in order to solve this minimization problem.
The smallest $q$ is limited to $0.5$, which corresponds to the minimum Hellinger
distance estimation \citep{Beran:1977}.
The case of $q < 0.5$ is not yet well-studied \citep{Qin+Priebe:2017}.

\section[Software]{Software} \label{sec:software}
\pkg{LqRT}, the package that implements all of the proposed tests above, is
available at \url{https://github.com/alyakin314/lqrt}.
In this section we present the three core functions of \pkg{LqRT}, which
correspond to the three versions of L$q$RT for testing the means of the normal 
distribution:
one-sample, two-sample paired and two-sample unpaired.

\subsection{One-sample test implementation}
The one-sample L$q$-likelihood-ratio-type test has the following function
signature in the \pkg{LqRT}
\begin{Code}
  lqrt.lqrtest_1samp(x, u,
  q = None, bootstrap=100,
  random_state=None)
\end{Code}
The first two positional arguments, \code{x} and \code{u}, represent the sample
observation and the expected value in null hypothesis in that order, similarly
to \fct{scipy.stats.ttest\_1samp}.
However, unlike \fct{scipy.stats.ttest\_1samp} and other $t$-tests of
\pkg{SciPy}, \fct{lqrt.lqrtest\_1samp} does not support testing multiple samples
or multiple hypotheses at once.
Thus, the sample \code{x} must be one-dimensional array-like, and \code{u} must
be a float.
The returned object is a named tuple of the test statistic $D_q$ and the
$p$-value, likewise for all of other testes in the \pkg{LqRT}.
This follows the convention of all the significance tests of \pkg{SciPy}.

The usage of the function is demonstrated in the example below.
In this example, we first test whether the mean of a sample from a normal
distribution is equal to its true value and then to a different value.
We do not reject the null hypothesis in the first case and do reject in the
second case.
\begin{CodeChunk}
  \begin{CodeInput}
    >>> import lqrt
    >>> import numpy as np
    >>> from scipy import stats
    >>>
    >>> np.random.seed(314)
    >>> rvs1 = stats.multivariate_normal.rvs(0, 1, 50)
    >>>
    >>> lqrt.lqrtest_1samp(rvs1, 0)
  \end{CodeInput}
  \begin{CodeOutput}
    Lqrtest_1sampResult(statistic=0.02388120731922072, pvalue=0.85)
  \end{CodeOutput}
  \begin{CodeInput}
    >>> lqrt.lqrtest_1samp(rvs1, 1)
  \end{CodeInput}
  \begin{CodeOutput}
    Lqrtest_1sampResult(statistic=35.13171144154751, pvalue=0.0)
  \end{CodeOutput}
\end{CodeChunk}
The optional argument \code{q} specifies the parameter $q$ of the
L$q$-likelihood.
The \code{q} typically should be within the interval [0.5, 1.0] and a lower
value is associated with a more robust test.
It can be specified manually or adaptively selected.
The latter happens if it is set to \code{None} or is left unspecified.
The adaptive selection procedures uses the trace of the empirical covariance
procedure, outlined in Section \ref{sec:select_q}.
An example below demonstrates the usage of two different manually provided
values for \code{q}, $0.9$ and $0.6$, as well as an adaptively selected one on
the same data, generated from a gross-error model.
\begin{CodeChunk}
  \begin{CodeInput}
    >>> rvs2 = np.concatenate([stats.multivariate_normal.rvs(0.32, 1, 45),
    ...                        stats.multivariate_normal.rvs(0.32, 50, 5)])
    >>>
    >>> lqrt.lqrtest_1samp(rvs2, 0, q=0.9)
  \end{CodeInput}
  \begin{CodeOutput}
    Lqrtest_1sampResult(statistic=2.239547159197258, pvalue=0.09)
  \end{CodeOutput}
  \begin{CodeInput}
    >>> lqrt.lqrtest_1samp(rvs2, 0, q=0.6)
  \end{CodeInput}
  \begin{CodeOutput}
    Lqrtest_1sampResult(statistic=3.4268748448623256, pvalue=0.02)
  \end{CodeOutput}
  \begin{CodeInput}
    >>> lqrt.lqrtest_1samp(rvs2, 0)
  \end{CodeInput}
  \begin{CodeOutput}
    Lqrtest_1sampResult(statistic=2.7337572196229587, pvalue=0.03)
  \end{CodeOutput}
\end{CodeChunk}
The $p$-value is obtained via a bootstrap procedure, described in Section
\ref{sec:critical_value}.
The number of samples in a bootstrap can be varied using the \code{boostrap}
argument to \fct{lqrt.lqrtest\_1samp}.
Increasing the number of samples increases the precision of the $p$-value, but
adds on computational work.
As a rough example, the three \fct{lqrt.lqrtest\_1samp} calls below took 0.3s,
1.5s and 15s, respectively.
\begin{CodeChunk}
  \begin{CodeInput}
    >>> lqrt.lqrtest_1samp(rvs1, 0, bootstrap=100)
  \end{CodeInput}
  \begin{CodeOutput}
    Lqrtest_1sampResult(statistic=0.02388120731922072, pvalue=0.85)
  \end{CodeOutput}
  \begin{CodeInput}
    >>> lqrt.lqrtest_1samp(rvs1, 0, bootstrap=1000)
  \end{CodeInput}
  \begin{CodeOutput}
    Lqrtest_1sampResult(statistic=0.02388120731922072, pvalue=0.875)
  \end{CodeOutput}
  \begin{CodeInput}
    >>> lqrt.lqrtest_1samp(rvs1, 0, bootstrap=10000)
  \end{CodeInput}
  \begin{CodeOutput}
    Lqrtest_1sampResult(statistic=0.02388120731922072, pvalue=0.8743)
  \end{CodeOutput}
\end{CodeChunk}
It should also be noted that the boostrapped resampling is random.
The argument \code{random_state} allows to seed the random number generator,
which allows reproducible results.

\subsection{Two-sample paired test implementation}
The two-sample L$q$RT L$q$-likelihood-ratio-type test has the function signature
\begin{Code}
  lqrt.lqrtest_rel(x_1, x_2,
  q=None, bootstrap=100,
  random_state=None)
\end{Code}
where \code{x_1} and \code{x_2} are two samples, which must be array-like,
one-dimensional and of equal size.
This function is a wrapper for the \code{lqrt.lqrtest_1samp}.
It extends the test from one-sample to a paired by subtracting one sample from
the other within the pairs and setting the null hypothesis mean, \code{u}, to
$0$.

We provide an example of the \fct{lqrt.lqrtest\_rel} usage.
First, we use the test on two samples from a normal distribution which actually
have identical population means:
\begin{CodeChunk}
  \begin{CodeInput}
    >>> import lqrt
    >>> from scipy import stats
    >>> import numpy as np
    >>> np.random.seed(314)
    >>>
    >>> rvs1 = stats.multivariate_normal.rvs(0, 1, 50)
    >>> rvs2 = stats.multivariate_normal.rvs(0, 1, 50)
    >>> 
    >>> lqrt.lqrtest_rel(rvs1, rvs2)
  \end{CodeInput}
  \begin{CodeOutput}
    Lqrtest_relResult(statistic=0.22769245832813567, pvalue=0.66)
  \end{CodeOutput}
\end{CodeChunk}
Now, we use the test on two samples drawn from the normal distributions with
different means:
\begin{CodeChunk}
  \begin{CodeInput}
    >>> rvs3 = stats.multivariate_normal.rvs(1, 1, 50)
    >>> lqrt.lqrtest_rel(rvs1, rvs3)
  \end{CodeInput}
  \begin{CodeOutput}
    Lqrtest_relResult(statistic=27.827284933987784, pvalue=0.0)
  \end{CodeOutput}
\end{CodeChunk}
The parameters \code{q}, \code{bootstrap} and \code{random_state} work
identically to the one-sample case, described in the previous section.


\subsection{Two-sample unpaired test implementation}
Lastly, we present a function that implements a two-sample unpaired L$q$RT
\begin{Code}
  lqrtest_ind(x_1, x_2, equal_var=True,
  q=None, bootstrap=100,
  random_state=None)
\end{Code}
The first two positional arguments, \code{x_1} and \code{x_2}, again, correspond
to the two test samples.
They must be array-like and one-dimensional, but need not be of the same size.
The test can be run with or without the equal population variance assumption.
This is done by varying the \code{equal_var} flag, similarly to the
\fct{scipy.stats.ttest\_ind}.
When set to \code{True}, the test corresponds to a more robust version of the
standard Student's $t$-test. When set to \code{False}, the test corresponds to a
more robust version of the Welch's $t$-test.
The default value of \code{equal_var} is \code{True}.
We present the examples of using the unpaired test, both with and without the
shared variance assumption.
\begin{CodeChunk}
  \begin{CodeInput}
    >>> import lqrt
    >>> from scipy import stats
    >>> import numpy as np
    >>> np.random.seed(314)
  \end{CodeInput}
\end{CodeChunk}
First, we generate two samples from a normal distribution that have different
sizes but are both centered around $0$.
We test whether their means are the same with and without the equal variance
assumption.
\begin{CodeChunk}
  \begin{CodeInput}
    >>> rvs1 = stats.multivariate_normal.rvs(0, 1, 50)
    >>> rvs2 = stats.multivariate_normal.rvs(0, 1, 70)
    >>> lqrt.lqrtest_ind(rvs1, rvs2)
  \end{CodeInput}
  \begin{CodeOutput}
    LqRtest_indResult(statistic=0.00046542438241203854, pvalue=0.99)
  \end{CodeOutput}
  \begin{CodeInput}
    >>> lqrt.lqrtest_ind(rvs1, rvs2, equal_var=False)
  \end{CodeInput}
  \begin{CodeOutput}
    LqRtest_indResult(statistic=0.00047040017227573117, pvalue=0.97)
  \end{CodeOutput}
\end{CodeChunk}
Now, we generate a new sample from a normal distribution which has a different
mean from the first two.
We then use the L$q$RT to test for the equivalence of means of the first sample
and the recently generated one.
We do so both with and without the equal variance assumption.
\begin{CodeChunk}
  \begin{CodeInput}
    >>> rvs3 = stats.multivariate_normal.rvs(1, 1, 70)
    >>> lqrt.lqrtest_ind(rvs1, rvs3)
  \end{CodeInput}
  \begin{CodeOutput}
    LqRtest_indResult(statistic=31.09168298440227, pvalue=0.0)
  \end{CodeOutput}
  \begin{CodeInput}
    >>> lqrt.lqrtest_ind(rvs1, rvs3, equal_var=False)
  \end{CodeInput}
  \begin{CodeOutput}
    LLqRtest_indResult(statistic=31.251454446588696, pvalue=0.0)
  \end{CodeOutput}
\end{CodeChunk}
The parameters \code{q}, \code{bootstrap} and \code{random_state}, once again,
work identically to the one-sample case.

\subsection{Real data example}
We also demonstrate a usage of the \pkg{LqRT} on the real data.
We use Breast Cancer Wisconsin (Diagnostic) Data Set as an example.
This data set can be easily import in \proglang{Python} using the
\pkg{scikit-learn} package \citep{scikit-learn}.
It is a copy of dataset located at the UC Irvine Machine Learning Repository ML
\citep{uci-datasets}.

The features are computed from a digitized image of a fine needle aspirate (FNA)
of a breast mass, and summarized as the mean, standard error, and worst
(largest) for each image, resulting in 30 total features per sample.
We only use the features' means, which corresponds to the first ten dimensions.
The dataset also includes binary labels corresponding to the tumor being
malignant or benign.
We present an example below in which we stratify the data on the true label and
use a two-sample unpaired test on each of the ten features.
\begin{CodeChunk}
  \begin{CodeInput}
    >>> import numpy as np
    >>> from sklearn.datasets import load_breast_cancer
    >>> np.random.seed(314)
    >>> 
    >>> X, y = load_breast_cancer(return_X_y=True)
    >>> X_negative = X[y==0]
    >>> X_positive = X[y==1]
    >>> features = 10
    >>> 
    >>> for i in range(features):
    ...     print(lqrt.lqrtest_ind(X_positive[:, i], X_negative[:, i],
    ...                            equal_var=False, bootstrap=1000))
  \end{CodeInput}
  \begin{CodeOutput}
    LqRtest_indResult(statistic=382.5469311314969, pvalue=0.0)
    LqRtest_indResult(statistic=77.89690998094738, pvalue=0.0)
    LqRtest_indResult(statistic=318.85934989217276, pvalue=0.0)
    LqRtest_indResult(statistic=207.32949298709445, pvalue=0.0)
    LqRtest_indResult(statistic=109.00139202641367, pvalue=0.0)
    LqRtest_indResult(statistic=384.764767407446, pvalue=0.0)
    LqRtest_indResult(statistic=758.937192422444, pvalue=0.0)
    LqRtest_indResult(statistic=1313.77261746955, pvalue=0.0)
    LqRtest_indResult(statistic=93.09648089232905, pvalue=0.0)
    LqRtest_indResult(statistic=0.0749898025942457, pvalue=0.84)
  \end{CodeOutput}
\end{CodeChunk}

\section[Experiments]{Experiments} \label{sec:experiments}
We compare the performance of the \pkg{LqRT} with other popular tests
implemented in \proglang{Python}, using the synthetically generated data.

\begin{table}[t!]
  \centering
  \begin{tabular}{lcccccccl}
    \hline
    Test                        & \multicolumn{2}{c}{Means (Size)}
    & \multicolumn{2}{c}{Means (Power)}
    & \multicolumn{3}{c}{Variances}
    & Hypothesis \\
    \hline
                                & $\mu_1$ & $\mu_2$ & $\mu_1$ & $\mu_2$
                                & $\sigma_1^2$ & $\sigma_2^2$ & $\tau$ \\
    \hline
    One-sample                  & $0$ &  -  & $0.34$ &  -
                                & $1$ &  -  & $50$   & $H_0: \mu_1 = 0$ \\
                                &     &     &        &
                                &     &     &        & $H_1: \mu_1 \neq 0$ \\
    \hline
    Two-sample paired           & $0$ & $0$ & $0$    & $0.50$
                                & $1$ & $1$ & $50$   & $H_0: \mu_1 = \mu_2$  \\
                                &     &     &        &
                                &     &     &        & $H_1: \mu_1 \neq \mu_2$ \\
    \hline
    Two-sample unpaired         & $0$ & $0$ & $0$    & $0.50$
                                & $1$ & $1$ & $50$   & $H_0: \mu_1 = \mu_2$  \\
    (equal variance assumption) &     &     &        &
                                &     &     &        & $H_1: \mu_1 \neq \mu_2$ \\
    \hline
    Two-sampled unapired        & $0$ & $0$ & $0$    & $0.50$
                                & $1$ & $0.01$ & $50$   & $H_0: \mu_1 = \mu_2$  \\
    (no equal variance assumption) &     &     &        &
                                &     &     &        & $H_1: \mu_1 \neq \mu_2$ \\
    \hline
  \end{tabular}
  \caption{\label{tab:parameters} Overview of the parameters used and the
    hypotheses tested in various set-ups for the gross error model simulation.}
\end{table}

In order to model the contamination in the data, we use a version of the gross
error model \citep{Huber:1964} in which both the underlying true distribution
and the anomalous values have a normal distribution.
The two distributions are centered at the same location, but the one
corresponding to the outliers is much wider.
Take $\epsilon$ to represent the probability of observing a gross error, or
an outlier.
This corresponds to the density of the form
\begin{align}
  g(x_i \vert \mu, \sigma^2, \tau^2, \epsilon)
  &= (1 - \epsilon) f(x_i \vert \mu, \sigma^2)
    + \epsilon f(x_i \vert \mu, \tau^2). \label{gross-error}
\end{align}
where $f$ is the density function of the normal distribution, $\epsilon < 0.5$
and $\sigma < \tau$. \citep[][page 357]{Bickel+Doksum}

The $\epsilon$ is typically not known a priori.
So, even though in all of the examples provided in this section the data is
generated according to the density in Equation \ref{gross-error}, the tests used
assume a regular one-dimensional two-parameter normal distribution.

\begin{table}[t!]
  \centering
  \begin{tabular}{ll}
    \hline
    Set-up                      & Tests used \\
    \hline
    One-sample                  & \fct{lqrt.lqrtest\_1samp} \\
                                & \fct{scipy.stats.ttest\_1samp} \\
                                & \fct{scipy.stats.wilcoxon} \\
                                & \fct{statsmodels.descriptivestats.sign\_test} \\
    \hline
    Two-sample paired           & \fct{lqrt.lqrtest\_rel} \\
                                & \fct{scipy.stats.ttest\_rel} \\
                                & \fct{scipy.stats.wilcoxon} \\
                                & \fct{statsmodels.descriptivestats.sign\_test} \\
    \hline
    Two-sample unpaired         & \fct[equal\_var=True]{lqrt.lqrtest\_ind} \\
    (equal variance assumption) & \fct[equal\_var=True]{scipy.stats.ttest\_ind} \\
                                & \fct{scipy.stats.ranksums} \\
    \hline
    Two-sampled unapired        & \fct[equal\_var=False]{lqrt.lqrtest\_ind} \\
    (no equal variance assumption) & \fct[equal\_var=False]{scipy.stats.ttest\_ind} \\
                                & \fct{scipy.stats.ranksums} \\
    \hline
  \end{tabular}
  \caption{\label{tab:testsused} Overview of the tests used in different
    set-ups in the gross error model simulation.}
\end{table}

We generate the samples from this model.
For all tests, the samples used had a size of $50$.
Other parameters used in the simulations to estimate the power against an
alternative are summarized in the Table \ref{tab:parameters}.
In the paired case we constrained the samples in the pair to either both have
gross errors, or both not be such.

\begin{figure}[t!]
  \centering
  \includegraphics[width=0.95\linewidth]{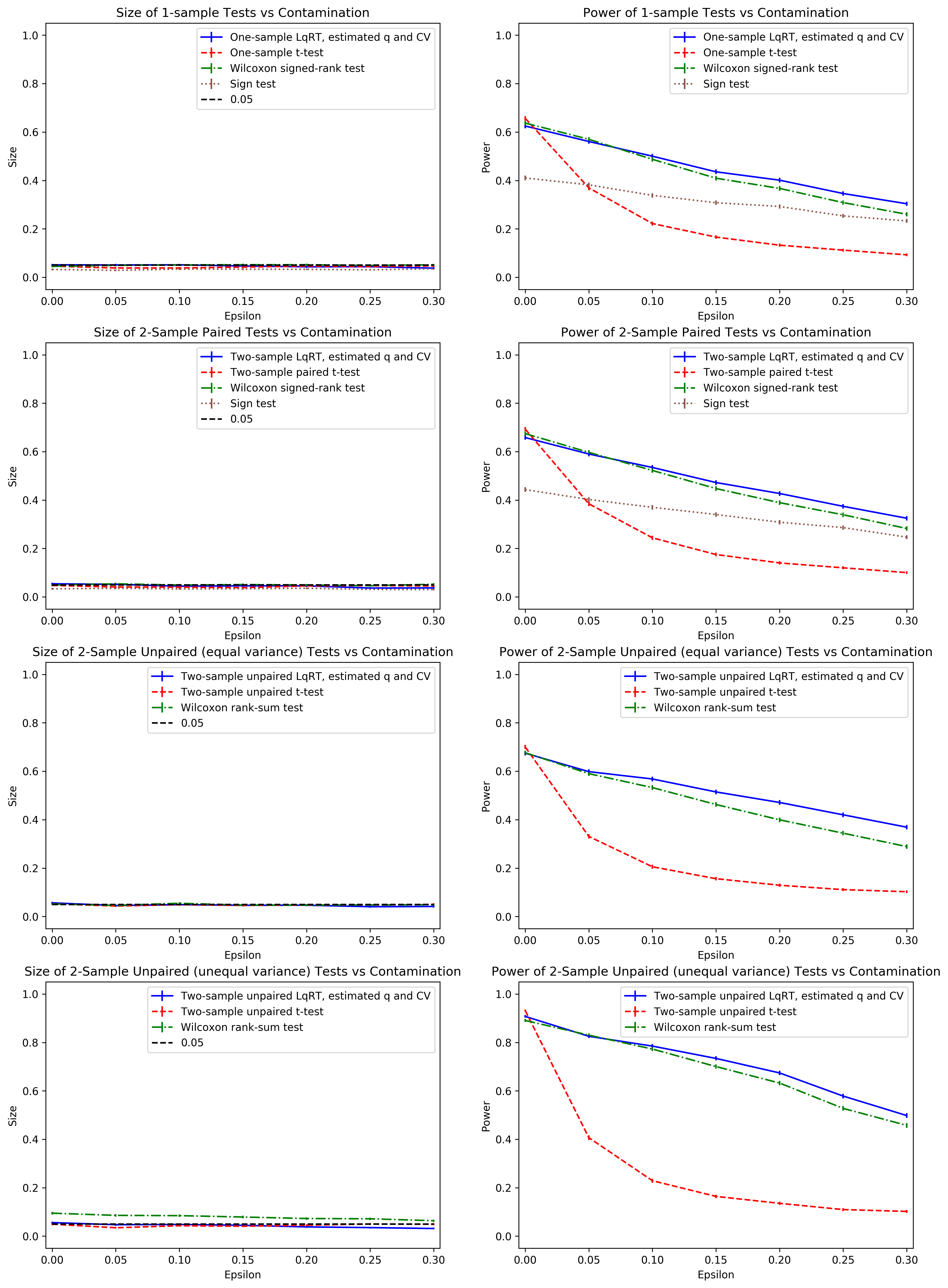}
  \caption{\label{fig:gemresults} Gross error model simulation
    results. Each datapoint is produced using 10000 repetitions. The error bars
    represent 95\% confidence interval.} 
\end{figure}

The performance of the \pkg{LqRT} was compared to that of other tests
implemented in \proglang{Python}.
The other tests used are listed in the Table \ref{tab:testsused}.
The results are presented in the Figure \ref{fig:gemresults}.
Across all of the set-ups, the tests implemnted in the \pkg{LqRT} are valid, in
the sense that the size is successfully controlled.
Furthermore, in all of the four cases, the power of the
L$q$-likelihood-ratio-type tests dominates all other tests for large enough
contamination.

\section[Conclusion]{Conclusion} \label{sec:conclusion}  
In this work we have summarized the existing methodology of the
L$q$-likelihood-ratio-type test and introduced a way to generalize it to the
two-sample unpaired Gaussian case.

We have also presented \proglang{Python} package, \pkg{LqRT}, offers an
interface to use all of the cases one-dimensional L$q$-likelihood-ratio-type
tests.
The package uses a syntax that closely resembles that of the statistical tests
in the commonly used \proglang{Python} packages, such as \pkg{SciPy}.
It has also been demonstrated on through simple examples that the tests
implemented in the \pkg{LqRT}, including the two-sample unpaired case, are valid
and perform as good or better than the competitors in terms of power when the
data is contaminated.

The possible directions for the future work, both on the theoretical results of
the L$q$-likelihood-ratio-type tests and the implementations, include extending
the Gaussian case to multiple dimensions, as well as implementing the tests for
the distribution other than Gaussian.
It is also possible to extend the package to the applications that commonly
include the assumption of normality, such as significance test of the
coefficients of the linear regression.


\bibliography{refs}

\begin{thebibliography}{22}
\newcommand{\enquote}[1]{``#1''}
\providecommand{\natexlab}[1]{#1}
\providecommand{\url}[1]{\texttt{#1}}
\providecommand{\urlprefix}{URL }
\expandafter\ifx\csname urlstyle\endcsname\relax
  \providecommand{\doi}[1]{doi:\discretionary{}{}{}#1}\else
  \providecommand{\doi}{doi:\discretionary{}{}{}\begingroup
  \urlstyle{rm}\Url}\fi
\providecommand{\eprint}[2][]{\url{#2}}

\bibitem[{Beran(1977)}]{Beran:1977}
Beran R (1977).
\newblock \enquote{Minimum Hellinger Distance Estimates for Parametric Models.}
\newblock \emph{Ann. Statist.}, \textbf{5}(3), 445--463.
\newblock \doi{10.1214/aos/1176343842}.
\newblock \urlprefix\url{https://doi.org/10.1214/aos/1176343842}.

\bibitem[{Bickel and Doksum(2006)}]{Bickel+Doksum}
Bickel P, Doksum K (2006).
\newblock \emph{Mathematical Statistics 2e}.
\newblock Pearson Education, Limited.
\newblock ISBN 9780131455924.
\newblock \urlprefix\url{https://books.google.com/books?id=U1CCkgEACAAJ}.

\bibitem[{Box and Cox(1964)}]{Box+Cox:1964}
Box GEP, Cox DR (1964).
\newblock \enquote{An Analysis of Transformations.}
\newblock \emph{Journal of the Royal Statistical Society. Series B
  (Methodological)}, \textbf{26}(2), 211--252.
\newblock ISSN 00359246.
\newblock \urlprefix\url{http://www.jstor.org/stable/2984418}.

\bibitem[{Conover(1999)}]{Conover:1999}
Conover W (1999).
\newblock \emph{Practical nonparametric statistics}.
\newblock Wiley series in probability and statistics: Applied probability and
  statistics. Wiley.
\newblock ISBN 9780471160687.
\newblock \urlprefix\url{https://books.google.com/books?id=dYEpAQAAMAAJ}.

\bibitem[{Dua and Graff(2017)}]{uci-datasets}
Dua D, Graff C (2017).
\newblock \enquote{{UCI} Machine Learning Repository.}
\newblock \urlprefix\url{http://archive.ics.uci.edu/ml}.

\bibitem[{Ferrari(2008)}]{Ferrari:2008}
Ferrari D (2008).
\newblock \enquote{Maximum Lq-likelihood Estimation.}
\newblock
  \urlprefix\url{https://conservancy.umn.edu/bitstream/handle/11299/60295/1/Ferrari_Davide\%20May\%202008.pdf}.

\bibitem[{Ferrari and Yang(2010)}]{Ferrari+Yang:2010}
Ferrari D, Yang Y (2010).
\newblock \enquote{Maximum Lq-likelihood Estimation.}
\newblock \emph{The Annals of Statistics}, \textbf{38}(2), 753--783.
\newblock \doi{10.1214/09-AOS687}.
\newblock \urlprefix\url{https://doi.org/10.1214/09-AOS687}.

\bibitem[{Grubbs(1969)}]{Grubbs:1969}
Grubbs FE (1969).
\newblock \enquote{Procedures for Detecting Outlying Observations in Samples.}
\newblock \emph{Technometrics}, \textbf{11}(1), 1--21.
\newblock \doi{10.1080/00401706.1969.10490657}.
\newblock
  \eprint{https://www.tandfonline.com/doi/pdf/10.1080/00401706.1969.10490657},
  \urlprefix\url{https://www.tandfonline.com/doi/abs/10.1080/00401706.1969.10490657}.

\bibitem[{Hampel \emph{et~al.}(2011)Hampel, Ronchetti, Rousseeuw, and
  Stahel}]{Hampel:2011}
Hampel F, Ronchetti E, Rousseeuw P, Stahel W (2011).
\newblock \emph{Robust Statistics: The Approach Based on Influence Functions}.
\newblock Wiley Series in Probability and Statistics. Wiley.
\newblock ISBN 9781118150689.
\newblock \urlprefix\url{https://books.google.com/books?id=XK3uhrVefXQC}.

\bibitem[{Heritier and Ronchetti(1994)}]{Heritier+Ronchetti:1994}
Heritier S, Ronchetti E (1994).
\newblock \enquote{Robust Bounded-Influence Tests in General Parametric
  Models.}
\newblock \emph{Journal of the American Statistical Association},
  \textbf{89}(427), 897--904.
\newblock ISSN 01621459.
\newblock \urlprefix\url{http://www.jstor.org/stable/2290914}.

\bibitem[{Hodges and Lehmann(1956)}]{Hodges+Lehmann:1956}
Hodges JL, Lehmann EL (1956).
\newblock \enquote{The Efficiency of Some Nonparametric Competitors of the
  $t$-Test.}
\newblock \emph{Ann. Math. Statist.}, \textbf{27}(2), 324--335.
\newblock \doi{10.1214/aoms/1177728261}.
\newblock \urlprefix\url{https://doi.org/10.1214/aoms/1177728261}.

\bibitem[{Huber(1964)}]{Huber:1964}
Huber PJ (1964).
\newblock \enquote{Robust Estimation of a Location Parameter.}
\newblock \emph{Ann. Math. Statist.}, \textbf{35}(1), 73--101.
\newblock \doi{10.1214/aoms/1177703732}.
\newblock \urlprefix\url{https://doi.org/10.1214/aoms/1177703732}.

\bibitem[{Huber(1965)}]{Huber:1965}
Huber PJ (1965).
\newblock \enquote{A Robust Version of the Probability Ratio Test.}
\newblock \emph{Ann. Math. Statist.}, \textbf{36}(6), 1753--1758.
\newblock \doi{10.1214/aoms/1177699803}.
\newblock \urlprefix\url{https://doi.org/10.1214/aoms/1177699803}.

\bibitem[{Jones \emph{et~al.}(2001--)Jones, Oliphant, Peterson
  \emph{et~al.}}]{SciPy}
Jones E, Oliphant T, Peterson P, \emph{et~al.} (2001--).
\newblock \enquote{\pkg{SciPy}: Open source scientific tools for
  \proglang{Python}.}
\newblock \urlprefix\url{http://www.scipy.org/}.

\bibitem[{Mann and Whitney(1947)}]{Mann+Whitney:1947}
Mann HB, Whitney DR (1947).
\newblock \enquote{On a Test of Whether one of Two Random Variables is
  Stochastically Larger than the Other.}
\newblock \emph{Ann. Math. Statist.}, \textbf{18}(1), 50--60.
\newblock \doi{10.1214/aoms/1177730491}.
\newblock \urlprefix\url{https://doi.org/10.1214/aoms/1177730491}.

\bibitem[{Murphy(2013)}]{Murphy:2013}
Murphy KP (2013).
\newblock \emph{Machine learning : a probabilistic perspective}.
\newblock MIT Press, Cambridge, Mass. [u.a.].
\newblock ISBN 9780262018029 0262018020.
\newblock
  \urlprefix\url{https://www.amazon.com/Machine-Learning-Probabilistic-Perspective-Computation/dp/0262018020/ref=sr_1_2?ie=UTF8\&qid=1336857747\&sr=8-2}.

\bibitem[{Pedregosa \emph{et~al.}(2011)Pedregosa, Varoquaux, Gramfort, Michel,
  Thirion, Grisel, Blondel, Prettenhofer, Weiss, Dubourg, Vanderplas, Passos,
  Cournapeau, Brucher, Perrot, and Duchesnay}]{scikit-learn}
Pedregosa F, Varoquaux G, Gramfort A, Michel V, Thirion B, Grisel O, Blondel M,
  Prettenhofer P, Weiss R, Dubourg V, Vanderplas J, Passos A, Cournapeau D,
  Brucher M, Perrot M, Duchesnay E (2011).
\newblock \enquote{\pkg{Scikit-learn}: Machine Learning in {P}ython.}
\newblock \emph{Journal of Machine Learning Research}, \textbf{12}, 2825--2830.

\bibitem[{{\proglang{Python} Software Foundation}(2001--)}]{Python}
{\proglang{Python} Software Foundation} (2001--).
\newblock \emph{The \proglang{Python} Language Reference}.
\newblock \urlprefix\url{https://docs.python.org/3/reference/}.

\bibitem[{Qin and Priebe(2017)}]{Qin+Priebe:2017}
Qin Y, Priebe CE (2017).
\newblock \enquote{Robust Hypothesis Testing via Lq-Likelihood.}
\newblock \emph{Statistica Sinica}, \textbf{27}, 1793--1813.

\bibitem[{Seabold and Perktold(2010)}]{Statsmodels}
Seabold S, Perktold J (2010).
\newblock \enquote{\pkg{Statsmodels}: Econometric and statistical modeling with
  python.}
\newblock In \emph{9th Python in Science Conference}.

\bibitem[{Tsallis(1994)}]{Tsallis:1994}
Tsallis C (1994).
\newblock \enquote{What are the numbers that experiments provide?}
\newblock \emph{Quimica Nova}, \textbf{17}, 468--471.
\newblock
  \urlprefix\url{http://submission.quimicanova.sbq.org.br/qn/qnol/1994/vol17n6/v17_n6_\%20(3).pdf}.

\bibitem[{Wilcoxon(1945)}]{Wilcoxon:1945}
Wilcoxon F (1945).
\newblock \enquote{Individual Comparisons by Ranking Methods.}
\newblock \emph{Biometrics Bulletin}, \textbf{1}(6), 80--83.
\newblock ISSN 00994987.
\newblock \urlprefix\url{http://www.jstor.org/stable/3001968}.

\end{thebibliography}


\newpage

\begin{appendix}

  \section{Other algorithms used} \label{app:algorithms}
  There are three other iterative re-weighting algorithms that are employed in
  the \pkg{LqRT} in addition to the Algorithm \ref{alg:reweighting_regular}.
  The pseudocode for all of them is provided below.

  The Algorithm \ref{alg:reweighting_known_mean} only optimizes the variance of
  a sample. It is used in the one-sample version of the test under the null
  hypothesis assumption.
  It is also implicitly used by the two-sample unpaired test, since it wraps
  around the one-sample.
  The Algorithms \ref{alg:reweighting_shared_variance} and
  \ref{alg:reweighting_shared_mean} are both used when some parameter is optimized
  jointly. The former estimates two different means but one shared variance and
  is used in the two-sample unpaired test with the shared variance assumption.
  The latter optimizes the mean jointly, but the variances separately and is
  used in the two-sample unpaired test without the shared variance assumption.

  \begin{algorithm}
    \caption{Iterative Re-weighting algorithm for an MLqE of a one sample from a
      one-dimensional normal with a known mean}
    \label{alg:reweighting_known_mean}
    \begin{algorithmic}[1]
      \Function{MLqE-Normal-Known-Mean}{$\boldsymbol{x}$, $u$, $q$}
      \State $n \gets $ \Call{Length}{$\boldsymbol{x}$}

      \Statex
      \Statex \quad \, \# Initialize Parameters with MLE
      \State $\hat{\sigma}^{2^{(0)}} \gets
      n^{-1} \sum_{i=1}^n (x_i - \mu)^2$

      \Statex
      \Statex \quad \, \# Iterative Re-weighting
      \For{$k=1, 2, \hdots \ until\ convergence$}

      \Statex
      \Statex \qquad \quad \# Update Weights
      \For{$i \gets 1, \hdots, n$}
      \State $w_i^{(s)} \gets \left( f \left( x_i \vert \hat{\mu}^{(s-1)},
          \hat{\sigma}^{2^{(s-1)}}\right) \right) ^ {1 - q}$
      \EndFor

      \Statex
      \Statex \qquad \quad \# Update Variance
      \State $\hat{\sigma}^{2^{(s)}} \gets \frac{\sum_{i=1}^n w_i (x_i - \mu)^2}
      {\sum_{i=1}^n w_i}$
      \EndFor

      \Statex
      \State \Return $\hat{\sigma}^{2^{(i)}}$
      \EndFunction
    \end{algorithmic}
  \end{algorithm}

  \begin{algorithm}
    \caption{Iterative Re-weighting algorithm for an MLqE of two samples from a
      one-dimensional normal with a shared variance constraint}
    \label{alg:reweighting_shared_variance}
    \begin{algorithmic}[1]
      \Function{MLqE-Normal-2sample-Equal-Variance}{$\boldsymbol{x}$,
        $\boldsymbol{y}$, $q$}
      \State $n \gets $ \Call{Length}{$\boldsymbol{x}$}
      \State $m \gets $ \Call{Length}{$\boldsymbol{y}$}
      
      \Statex
      \Statex \quad \, \# Initialize Parameters with MLE
      \State $\hat{\mu}_x^{(0)} \gets n^{-1} \sum_{i=1}^n x_i$
      \State $\hat{\mu}_y^{(0)} \gets m^{-1} \sum_{i=1}^n y_i$
      \State $\hat{\sigma}^{2^{(0)}} \gets
      (n + m)^{-1} \left(\sum_{i=1}^n (x_i - \hat{\mu}_x^{(0)})^2
        + \sum_{h=1}^m (y_j - \hat{\mu}_y^{(0)})^2 \right)$

      \Statex
      \Statex \quad \, \# Iterative Re-weighting
      \For{$k=1, 2, \hdots \ until\ convergence$}
      \Statex
      \Statex \qquad \quad \# Update Weights
      \For{$i \gets 1, \hdots, n$}
      \State $w_{x,i}^{(s)} \gets \left( f \left( x_i \vert \hat{\mu}_x^{(s-1)},
          \hat{\sigma}^{2^{(s-1)}}\right) \right) ^ {1 - q}$
      \EndFor
      \For{$j \gets 1, \hdots, n$}
      \State $w_{y,j}^{(s)} \gets \left( f \left( y_j \vert \hat{\mu}_y^{(s-1)},
          \hat{\sigma}^{2^{(s-1)}}\right) \right) ^ {1 - q}$
      \EndFor

      \Statex
      \Statex \qquad \quad \# Update Parameters
      \State $\hat{\mu}_x^{(s)} \gets \frac{\sum_{i=1}^n x_i}{\sum_{i=1}^n w_{x,i}}$
      \State $\hat{\mu}_y^{(s)} \gets \frac{\sum_{j=1}^m y_j}{\sum_{j=1}^n w_{y,j}}$
      \State $\hat{\sigma}^{2^{(s)}} \gets
      \frac{\sum_{i=1}^n w_{x,i}^{(s)} (x_i - \hat{\mu}_x^{(s)})^2
        + \sum_{j=1}^n w_{x,i}^{(s)} (y_j - \hat{\mu}_y^{(s)})^2
      }{\sum_{i=1}^n w_{x,i}^{(s)} + \sum_{j=1}^n w_{y,j}^{(s)}}$

      \Statex
      \Statex \qquad \quad \# Clip Variance
      \If{$\hat{\sigma}^{2^{(s)}} < \epsilon$} 
      \State $\hat{\sigma}^{2^{(s)}} \gets \epsilon$
      \EndIf
      \EndFor

      \Statex
      \State \Return $\hat{\mu}_x^{(s)}, \hat{\mu}_y^{(s)}, \hat{\sigma}^{2^{(i)}}$
      \EndFunction
    \end{algorithmic}
  \end{algorithm}

  \begin{algorithm}
    \caption{Iterative Re-weighting algorithm for an MLqE of two samples from a
      one-dimensional normal with a shared mean constraint.
    }
    \label{alg:reweighting_shared_mean}
    \begin{algorithmic}[1]
      \Function{MLqE-Normal-2sample-Equal-Mean}{$\boldsymbol{x}$,
        $\boldsymbol{y}$, $q$}
      \State $n \gets $ \Call{Length}{$\boldsymbol{x}$}
      \State $m \gets $ \Call{Length}{$\boldsymbol{y}$}
      
      \Statex
      \Statex \quad \, \# Initialize Parameters with MLE
      \State $\hat{\mu}^{(0)} \gets (n + m)^{-1}
      \left(\sum_{i=1}^n x_i + \sum_{i=1}^m y_i\right)$
      \State $\hat{\sigma}_x^{2^{(0)}} \gets
      n^{-1} \sum_{i=1}^n (x_i - \hat{\mu}^{(0)})^2$
      \State $\hat{\sigma}_y^{2^{(0)}} \gets
      m^{-1} \sum_{j=1}^m (y_j - \hat{\mu}^{(0)})^2$

      \Statex
      \Statex \quad \, \# Iterative Re-weighting
      \For{$k=1, 2, \hdots \ until\ convergence$}

      \Statex
      \Statex \qquad \quad \# Update Weights
      \For{$i \gets 1, \hdots, n$}
      \State $w_{x,i}^{(s)} \gets \left( f \left( x_i \vert \hat{\mu}^{(s-1)},
          \hat{\sigma}_x^{2^{(s-1)}}\right) \right) ^ {1 - q}$
      \EndFor
      \For{$j \gets 1, \hdots, n$}
      \State $w_{y,j}^{(s)} \gets \left( f \left( y_j \vert \hat{\mu}^{(s-1)},
          \hat{\sigma}_y^{2^{(s-1)}}\right) \right) ^ {1 - q}$
      \EndFor
      
      \Statex
      \Statex \qquad \quad \# Update Parameters
      \State $\hat{\mu}^{(s)} \gets
      \frac{\sum_{i=1}^n w_{x,i}^{(s)} x_i + \sum_{j=1}^m w_{y,j}^{(s)} y_i}
      {\sum_{i=1}^n w_{x,i}^{(s)} + \sum_{j=1}^m w_{y,j}^{(s)}}$
      \State $\hat{\sigma}_x^{2^{(s)}} \gets \frac{\sum_{i=1}^n w_{x,j} (x_i -
        \hat{\mu}^{(s)})^2}{\sum_{i=1}^n w_{x,i}}$
      \State $\hat{\sigma}_y^{2^{(s)}} \gets \frac{\sum_{j=1}^m w_{y,j} (y_j -
        \hat{\mu}^{(s)})^2}{\sum_{j=1}^m w_{y,j}}$
      
      \Statex
      \Statex \qquad \quad \# Clip Variances
      \If{$\hat{\sigma}_x^{2^{(s)}} < \epsilon$}
      \State $\hat{\sigma}_x^{2^{(s)}} \gets \epsilon$
      \EndIf
      \If{$\hat{\sigma}_y^{2^{(s)}} < \epsilon$}
      \State $\hat{\sigma}_y^{2^{(s)}} \gets \epsilon$
      \EndIf

      \Statex
      \State \Return $\hat{\mu}^{(s)}, \hat{\sigma}_x^{2^{(s)}},
      \hat{\sigma}_y^{2^{(s)}}$
      \EndFor
      \EndFunction
    \end{algorithmic}
  \end{algorithm}

\end{appendix}
\clearpage

\end{document}